\documentclass[11pt]{article}
\usepackage[totalwidth=385pt,totalheight=580pt]{geometry}
\usepackage{amsmath,amssymb,epsf,graphics,psfrag}
\usepackage{graphicx}
\usepackage{psfrag}

\newcommand{\ket}{\rangle}
\newcommand{\sket}[1]{| #1\rangle}
\newcommand{\sbra}[1]{\langle #1 |}
\newcommand{\bra}{\langle}

\newcommand\eq{\begin{equation}}
\newcommand\en{\end{equation}}
\newcommand\bea{\begin{eqnarray}}
\newcommand\eea{\end{eqnarray}}
\newcommand\nn{\nonumber}
\newcommand\ba{\(\begin{array}}
\newcommand\ea{\end{array}\)}
\newcommand{\resection}[1]{\setcounter{equation}{0}\section{#1}}

\newcommand{\D}{{\cal D}}
\newcommand{\Z}{{\mathbb Z}}
\newcommand{\N}{{\mathbb N}}
\newcommand{\R}{{\mathbb R}}
\newcommand{\os}{{\mathbb O}}

\newcommand\tr{{\rm tr}\,}
\begin{document}

\begin{titlepage}
\vskip 0.5cm
\begin{flushright}

Today's date: \today
\end{flushright}
\vskip .7cm
\begin{center}
{\Large{\bf  Entanglement entropy and the complex plane of replicas}}
\end{center}
\vskip 0.8cm \centerline{Ferdinando  Gliozzi$^3$ and Luca Tagliacozzo$^1$} 
\vskip 0.9cm

\vskip 0.3cm \centerline{${}^{3}$\sl\small Dip.\ di Fisica Teorica
and INFN, Universit\`a di Torino,} \centerline{\sl\small Via P.\
Giuria 1, 10125 Torino, Italy}

\vskip 0.3cm \centerline{${}^{1}$\sl\small School of Physical Sciences, the University of Queensland, QLD 4072, Australia}

\vskip 1.25cm
\begin{abstract}
\noindent

The entanglement entropy of a subsystem $A$ of a quantum system is
expressed, in the replica method, through analytic continuation with
respect to $n$ of the trace of the ${\rm n}^{th}$ power of the reduced density
matrix $\tr\rho_A^n$. We study the analytic properties of this quantity
as a function of $n$ in some  quantum critical Ising-like  models in 1+1 and 
2+1 dimensions. Although we find  no true singularities for $n>0$, there is 
a threshold value of $n$ close to 2  which separates two very different 
`phases'. The region with larger $n$ is characterised 
by rapidly convergent Taylor expansions and is very smooth. The region 
with smaller $n$  has a very rich and varied structure in the  complex $n$ 
plane and is characterised by 
Taylor coefficients which instead of being monotone 
decreasing, have a maximum growing with the size of the subsystem.  
Finite truncations of the Taylor expansion in this region lead to  increasingly poor 
approximations of  $\tr\,\rho_A^n$. The computation of the entanglement entropy from the knowledge of
$\tr\rho^n_A$ for positive integer $n$ becomes  extremely difficult  particularly  in spatial dimensions larger than one, 
where one cannot use conformal field theory as a guidance in the extrapolations to $n=1$.

\end{abstract}
\end{titlepage}
\setcounter{footnote}{0}

\def\thefootnote{\fnsymbol{footnote}}

\resection{Introduction}
The entanglement entropy has become a privileged measure of bipartite entanglement for pure states, that is of the entanglement between a subregion of the system and the rest of it. 
In order to compute it, one divides the system in the  state $\sket{\Psi}$, into two complementary subsystems $A$ and $B$. The reduced density matrix of the region $A$ is  obtained by  tracing out the degrees of freedom inside the region $B$, $ \rho_A= \tr_{B} \sket{\Psi}\sbra{\Psi}$. The correlations between $A$ and $B$ give rise to  a mixed state for $\rho_A$ whose  von Neumann entropy $S_A\equiv -\tr\rho_A\log \rho_A$ is the entanglement entropy of the region $A$ (see e.g. \cite{rev} for reviews). 
 Other  quantities such as the R\'enyi ($R(\alpha)$) \cite{reny} and the Tsallis entropies
($T(\alpha)$) \cite{tsallis} 
\eq
R(\alpha)=\frac{\log\tr\rho_A^\alpha}{1-\alpha}~;~~~~T(\alpha)=\frac{\tr\rho^\alpha-1}{1-\alpha}~,~~~\alpha>0~,
\label{rets}
\en
are strictly related to the entanglement entropy of the block $A$. Indeed both of them coincide with  it in the  limit $\alpha \to 1$. 

Studying directly the entanglement entropy in the ground state  $\sket{\Psi}$  of a chosen Hamiltonian  requires the ability to obtain such ground state.  The complexity of this, in general, grows   exponentially with the size of the system (recently new methods have been proposed to approximate the ground state  of two dimensional systems that only scale polynomially with the system size \cite{mera,peps}). 

 However, when $\alpha$  in Eq. (\ref{rets}) is  a positive integer  $n$,   the Tsallis and  R\'enyi entropies  can be computed in the context of 
quantum field theory (QFT), without the explicit knowledge of the ground state. The procedure is known as  the ``replica method'' \cite{hlw,cc} and is based on the fact that the expression $tr\rho_A^n$ can be written as a partition function of the model on a $n-$sheeted Riemann surface (or its generalization in $d+1$ dimensions), divided by $Z^n$, where $Z$ is the partition function of the original system  . 
The entanglement entropy is obtained from it through  an analytical  continuation from positive integer $n$ to real  $\alpha$ using  $S_A=-\lim_{\alpha\to1}\frac\partial{\partial \alpha}\tr\rho_A^\alpha$.
Even if  all values of $\tr\rho^n_A$ were provided for $n\in\N$, analytic 
continuation to real $\alpha$ might not be  uniquely determined, as examples on disordered systems show \cite{hp}.  It is therefore desirable
to have some information on the analytical properties of
$\tr\rho^\alpha_A$. 

In the case of $1+1$ conformal  field theories (CFT)  the behaviour of the 
$T(n)$ when $A$  is composed by a single  interval of length  $\ell$ and is a part of an infinite gapless system is known to be
 \cite{hlw,cc}
\eq
\tr\rho_\ell^n=r_n\,\left(\frac a\ell\right)^{c(n-1/n)/6}~~.
\label{cc}
\en
Here $a$ is an ultraviolet (UV) cutoff, $c$ is the central charge of the associated CFT and $r_n$ is a non-universal function of $n$ which is known only in some special cases. In some cases even the full analytic continuation  has been performed \cite{cardy-twist} allowing one to  extract also the non universal additive corrections to the scaling  of the entanglement entropy .

  In $d>1$, however, CFT results are not available and one thus needs to carefully study the analytical properties of  $\tr\rho^\alpha_A$ before trying to perform any analytic continuations of it. 

The purpose of this work is  to perform such a  study by exploring   
the analytic properties of  $\tr\rho_A^\alpha$ as a function of  $\alpha$ in the whole complex plane for a class of quantum Ising-like  critical  models in both  1+1 and 2+1 dimensions.
 By doing this  we uncover a highly non-trivial analytical behaviour 
of $\tr\rho_A^\alpha$ for $\alpha$ real in the interval $0<\alpha<2$.
 In all the models considered, we find a threshold value $n_c$  in the interval $1<n_c<2$ which divides $\tr\rho^\alpha_A$ into two `phases'.
 In the region with $\alpha\ge n_c$ the quantity $\tr\rho_A^\alpha $ behaves very smoothly as a function of $\alpha$; by expanding it in a  Taylor series centred around  any of the $\alpha_o > n_c$, we extract alternating series with monotone 
decreasing coefficients, then, any truncation
$\sum_{k=0}^{n-1}c_k(\alpha-\alpha_o)^k$ can be used to approximate   $\tr\rho_A^\alpha$,
the error of this approximation being less than 
$\vert c_{n} (\alpha-\alpha_o)^n\vert$. For this  reason we call the  
region  $\alpha\ge n_c$ `perturbative' region.

\begin{figure}
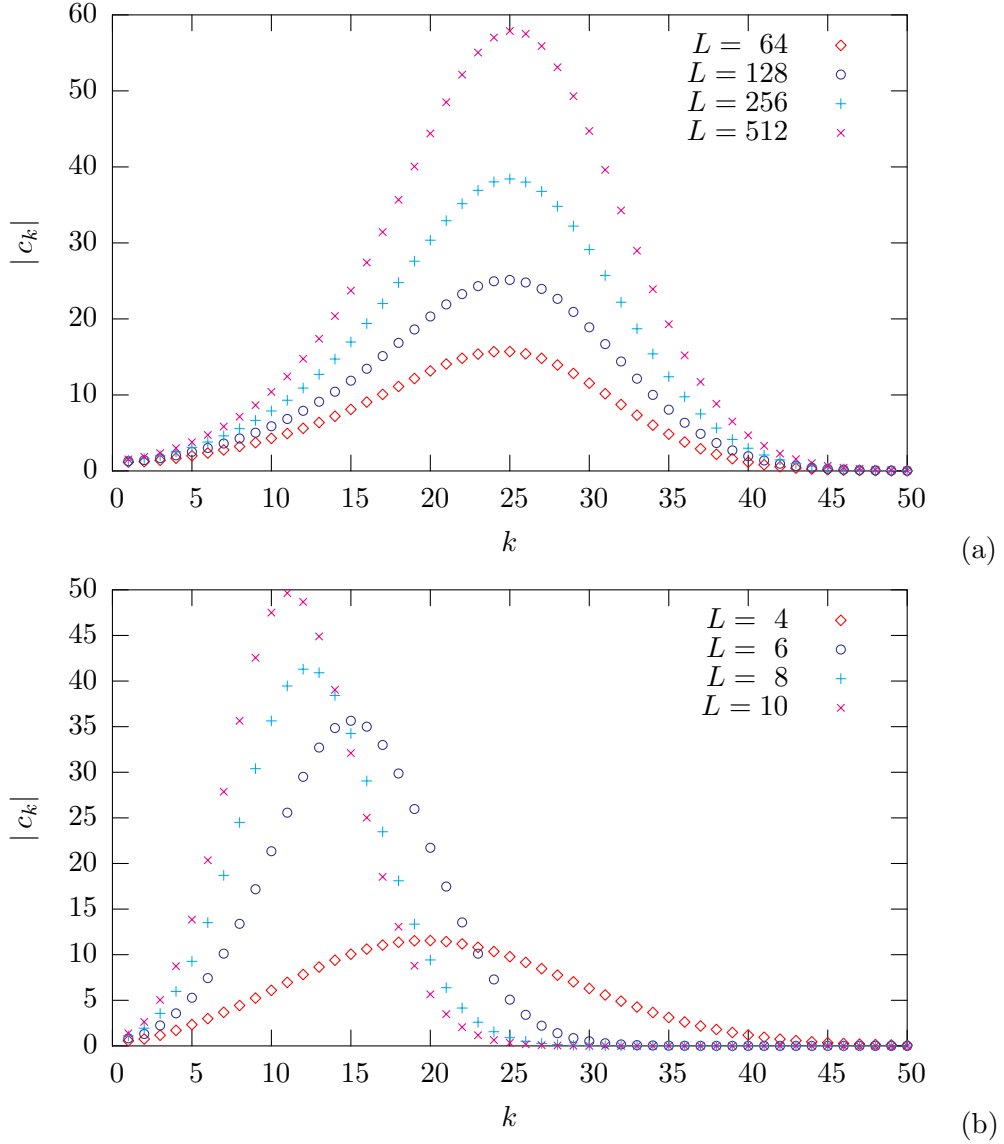

\begin{center}
\input Refig/tay1D \\
\input Refig/tay2D
\end{center}
\caption{Plot of the moduli of the coefficients of the Taylor expansion of
	of $\tr\rho_A^\alpha$ about $\alpha=1$ for (a) the 1D Ising model in transverse field at the critical point and (b) the 2D Ising model defined on a torus of $L\times L$. In (a) the block $A$ consists of $L$ adjacent spins of the 
infinite chain with $L=64,128,256,512$. The spectrum of the reduced matrix 
$\rho_A$ is obtained applying  the method described in Ref.\cite{vlrk}. 
In (b) the subsystem $A$ is half a torus $L\times L$ with $L=4,6,8,10$. The 
spectrum is obtained by using the method described in Ref. \cite{tev}. 
In both dimensions the height of the peaks of the Taylor coefficients increases with L.}
\label{Figure:1}
\end{figure}

On the contrary, in the other `phase', the Taylor expansions of  $\tr\rho_A^\alpha $ with $0\le\alpha_o<n_c$  has  coefficients whose moduli  are no longer monotone  decreasing. They show a characteristic peak 
whose height  increases  with the size of  $A$  (see for instance Fig.\ref{Figure:1}). In this regime  any truncation of the Taylor series becomes an  increasingly  poor approximation of  $tr\,\rho_A^n$ as the size of $A$ grows. Thus we  name  the region 
$0\le\alpha<n_c$  `non-perturbative' region.

We shall see that the threshold value $n_c$ coincides with the point where
the expression
 \mbox{$\frac{{\rm d}^2  \tr\rho_A^\alpha  }{{\rm d}\alpha^2} +\frac13 \frac{{\rm d}^3\tr\rho_A^\alpha  }{{\rm d}\alpha^3}$} changes sign, that is 
\eq
 \left(\frac{{\rm d}^2}{{\rm d}\alpha^2} + 
\frac13 \frac{{\rm d}^3}{{\rm d}\alpha^3}\right)  \tr\rho_A^{\alpha}\vert_{\alpha=n_c}=0~.
\label{order}
\en
In the models we have considered it turns out that the two different `phases' 
are separated by a smooth cross-over in both  $d=1$ and $d=2$. However
we cannot exclude that, in higher dimensions, this cross-over turns into a real phase transition. In fact  it has been recently observed in Ref.   \cite{sach}  that the R\'enyi entropies in the $O(N)$ model undergo a phase transition as a function of $\alpha$ for $d$ close to 3. This happens in the same range of $\alpha$  in which   the  models we analyse develop the cross-over between the perturbative and the non-perturbative regions.

The above scenario has important consequences when  trying to extract the entanglement entropy  using the replica trick. 
On one side the absence of a real phase transition makes the analytic 
continuation (it involves) possible. 
On the other side, the different behaviour of the two 'phases' implies that 
finding it can be  really a non trivial task.

In fact, it is easy to find a simple function $f(n)$ 
  which accurately  reproduces   $\tr\rho_A^n$  for integer $n$. However, if we 
make the `obvious' extension to real $n$, by assuming  
that this functional form is also valid for real $n$, we obtain an incorrect value of the entanglement entropy. 

By exemplifying this we also have the opportunity to compare  two different methods. One is  based on a recent 
algorithm  proposed in Ref. \cite{tev}
which allows us to accurately evaluate the reduced density matrix of a 
two-dimensional quantum system 
defined on a lattice of small size (in this paper we generally refer to  
this method  as the Hamiltonian approach).   
The other one is based on the replica method and  applies a simple
technique described in Ref. \cite{cg}  to directly measure $\tr\rho_A^n$  
as the vacuum expectation  value of a suitable observable in a  Monte Carlo 
simulation, in  any Euclidean lattice field theory.

The two methods agree on the Tsallis entropies $T(n)$  for a  wide 
range of  integer $n\ge2$, once  the UV cutoff of the two approaches
 are matched through 
a suitable conversion  factor (this by itself is a non trivial result that   
provides a strong confirmation of the validity of both  approaches).
However there is a mismatch  in the results for the  entanglement entropy. Its value directly computed in the  Hamiltonian approach (the correct one) is different from the one obtained  in the Monte Carlo approach  from the naive continuation of  the behaviour of the Tsallis entropies for integer $n$  to  real $\alpha$. This is a consequence of the intricate landscape of $\tr\rho_A^n$ close 
to $n=1$. 
The  scenario  we have outlined is common to all critical quantum systems 
in one and two dimensions we have analysed.

The contents of this paper  are organised in several sections. 
In  section \ref{sec:model} we describe the 
quantum models we want to study  with different  numerical methods and, where available,  analytical results.
In section \ref{mc}  we describe both  the  Monte Carlo and the variational  
techniques  we use to analyse quantum systems in 2+1 dimensions. 
Section \ref{sec:tay} and \ref{sing}  present the main results of this paper. 
In Section \ref{sec:tay}  we outline the presence of two `phases' 
for $\tr\rho_A^\alpha$  by considering the  features of its Taylor expansions. In section \ref{sing} we look for the Lee-Yang zeros and other possible singularities 
in the complex plane of $\alpha$ and we find no singularity related to the 
transition between these two regions. We unveil the presence of a smooth 
cross-over between the two regions.
In Section \ref{ent_ent} we discuss  the effects  of this  scenario on 
entanglement entropy calculations. There we also present the  
comparison between  the numerical results obtained with Monte Carlo and 
variational techniques. 
 We conclude  with a discussion of the results in Section \ref{sec:con}. 

\resection{The models} 
\label{sec:model}
 We analyzed the reduced density matrix $\tr\rho^n_A$  of the ground state near and at the quantum  critical point in a series of one-dimensional 
spin $\frac12$ chain systems and their two-dimensional generalizations.
In particular, we considered the XY Hamiltonian
\eq
H_{\rm XY}=-\sum_{\bra i,j\ket}\left[(1+\gamma)\sigma_i^x\sigma_j^x+(1-\gamma)\sigma_i^y\sigma_j^y
\right]-2\,h\sum_i\sigma_i^z~, 
\label{XY}
\en
where $\bra i,j\ket$ denote first neighbor sites, $\sigma_i^\alpha$ are Pauli matrices and 
$0\le\gamma\le1$, $h\ge0$. The subsystem $A$ is a block of neighboring nodes.

We applied mainly the numerical methods described in \cite{vlrk,tev}, but we used also analytical results. Analytical calculations exist for the critical XX chain $(\gamma=0)$ \cite{jk}, for the 
non-critical XY chain in a field \cite{ijk} as well as for the 
Heisenberg chain \cite{ps}.

Simplifications arise when mapping the quantum chains into classical two-dimensional classical spin systems \cite{ni}. The quantum XY chain can be mapped, at least for a part of the parameter range, into a classical  Ising model on a triangular lattice \cite{ip}. In this way simple formulas can be easily obtained. For 
instance, if the subsystem $A$ is a block of spins of size $L$ much larger than the correlation 
length $\xi$, embedded into an infinite chain with periodic boundary conditions at zero temperature
in the ordered phase ((i.e. $h<1$) the reduced matrix is independent of $L$ and can be written in the simple form \cite{fik}
\eq
\tr\rho^\alpha_L=
2\left(\frac{\prod_{n=1}^\infty(1+q^{n\,\alpha})}{\prod_{n=1}^\infty(1+q^n)^\alpha}\right)^2~,\,q=\exp(-2\epsilon)~,
\label{rhoxy}
\en    
where the parameter $\epsilon$  which sets the scale is a known function of $\gamma$ and $h$ 
\cite{ip,fik}. The factor of two in front of (\ref{rhoxy}) arises from the asymptotic degeneracy of the ground state in the broken $\Z_2$ symmetry for $h<1$. The exponent of two in (\ref{rhoxy}) 
reflects instead the fact that the segment $L$ in a chain with periodic boundary conditions has 
two points of contact, i.e. two interfaces with the rest of the system.    

We are interested in the critical limit $\epsilon\to0$ where the infinite products (\ref{rhoxy})
become slowly convergent, therefore we rewrite (\ref{rhoxy}) in a more suitable form using the identity
\eq
\prod_{n=1}^\infty(1+q^{n\,\alpha})=\frac{\eta(2\alpha\tau)}{\eta(\alpha\tau)}\exp(-i\pi\alpha\tau/12)~,\,\tau=i\frac\epsilon\pi~,
\en
thus
\eq
\tr\rho^\alpha_L=2\left(\frac{\eta(2\alpha\tau)\,\,\eta(\tau)^\alpha}
{ \eta(\alpha\tau)\,\,\eta(2\tau)^\alpha }\right)^2~.
\en
 we introduced the Dedekind $\eta$ function
\eq
\eta(\tau)=q^{\frac1{24}}\prod_{n=1}^\infty(1-q^n)
\label{eta}
\en
because it has  a simple behaviour under the modular transformation $\tau\to-1/\tau$, namely
\eq
\eta(\tau)=\frac1{\sqrt{-i\tau}}\eta(-\frac1\tau)~,
\en 
and this leads to rapidly converging products. Precisely we have
\eq
\tr\rho^\alpha_L=\exp\left(\frac{\pi^2}{12\epsilon}(\frac1\alpha-\alpha)\right)\,2^\alpha\,
\left(\frac{\prod_{n=1}^\infty\left(1+\exp(-\frac{\pi^2}\epsilon n)\right)^\alpha}
{\prod_{n=1}^\infty\left(1+\exp(-\frac{\pi^2}\epsilon \frac n\alpha)\right)}\right)^2~.
\label{rhoXY}
\en  
Since $\frac{\pi^2}\epsilon\simeq\log\xi$, the first exponential tells us, according to 
\cite{hlw,cc}, that the associated  CFT in the critical limit is the free fermion model with central charge $c=\frac12$ as expected, however the exact expression (\ref{rhoXY}) allows to discuss 
the analytical structure of the reduced density matrix in the whole complex replica plane. 
This discussion is postponed to Section \ref{sing}.

For the other critical models, to our knowledge there are no analytical results available for the reduced density matrices, and  we have to rely on  a set of different numerical techniques. For the generic one dimensional critical  XY chains  we  use the expression of the reduced density matrices of an interval  in terms of the free fermionic modes described in Ref. \cite{vlrk} (for more recent results  see e. g.  Ref. \cite{kore_its}) and numerically diagonalise it. In the case of finite chains at the critical point we use the one dimensional version  of the variational technique described in Ref. \cite{tev} that employs a Tree Tensor Network as an Ansatz for the ground state of the system.

 In two dimensions we analyse the Ising model in a transverse field  and 
the XX model on finite tori  with both the Monte Carlo technique described 
in  Ref. \cite{cg} and the  variational technique of Ref. \cite{tev} that 
uses again a Tree Tensor Network. In 2D the subsystem $A$ is one half 
of a  torus bounded by two parallel straight lines.
\resection{Monte Carlo simulations and variational algorithms}
\label{mc}
In QFT the quantity $\tr\rho^n_A$  can be
 written as the 
vacuum expectation value of a suitable observable $\os$ defined on a larger system composed of $n$ decoupled copies of the original system. 

The canonical partition function $Z=\tr e^{-\beta H}$ of  a 
$d-$dimensional quantum system at inverse temperature $\beta$ can be computed 
  in a standard way  by doing the Euclidean  functional integral in a 
$d+1$-dimensional  hyper-cubic lattice $\Lambda=\{\vec{x},\tau\}$  $(x_i,\tau\in\Z)$ over fields $\phi(x)\equiv
\phi(\vec{x},\tau)$ periodic under $\tau\to\tau+\beta$. Therefore the system 
composed of $n$ independent replicas of the original system is described by 
the $n-$th power of $Z$: 
\eq
Z^n=\int\prod_{k=1}^n\D[\phi_k] e^{-\sum_{k=1}^nS[\phi_k]}~,
\label{nz}
\en   
where $\phi_k$ is a field configuration associated to the $k-$th
copy and $S[\phi]$ is the Euclidean action.

In the replica approach to entanglement the subsystem $A$ establishes 
a process of transferring information among the $n$ replicas through a 
specific  coupling: the lattice links coming out of nodes of 
$A$ and directed into the (Euclidean) 
time direction $\tau$ cyclically 
connect the copy $k$ with the copy $k+1$   \cite{hlw,cc}. Let us  denote with 
$ S_A[\phi_1,\phi_2,\dots,\phi_n]$ the corresponding coupled action. 
It is  easy  to see that the quantity 
\eq
\os=e^{-\left(S_A^{(n)}[\phi_1,\phi_2,\dots,\phi_n]-\sum_{k=1}^nS[\phi_k]
\right)}~,
\en
has the desired property. In fact
its vacuum expectation value in the system of $n$ independent copies of the
original system is
\eq
\bra\os\ket_n=\frac{\int\prod_{k=1}^n\D\phi_k\,\os\,e^{-S[\phi_k]}}{Z^n}
=\frac{\int\prod_{k=1}^n\D\phi_k\,e^{-S_A^{(n)}}}{Z^n}
=\frac{Z_n(A)}{Z^n}=\tr\rho_A^n~.
\label{obs}
\en
$Z_n(A)$ is the partition function of the system in the $n$ coupled
replicas.

This method can be simply implemented in Monte Carlo simulations. 
Here we apply it to the case in which the system is a periodic
 lattices of size $L^2\times 8L$, where the longer direction is  the Euclidean 
time. We verified that the Euclidean time is large enough to  ensure the 
absence of measurable finite temperature effects. 
The Monte Carlo simulations were performed with a particularly efficient implementation for the Ising and 
$q-$ state Potts model described in \cite{cg} and with  the standard 
action of the isotropic Ising model on a cubic lattice 
\eq
S=-\sum_{\bra i,j\ket}S_i\,S_j~;~~S_i=\pm1~.
\en
All simulations were made at the critical coupling $\beta\simeq0.221652$.

With this  method we compute directly the Tsallis entropies for integer $n$. A discussion of the results we obtained is postponed to Section \ref{ent_ent}.

Using the techniques of Ref. \cite{tev} we also extract an accurate variational approximation to the ground state with a Tree Tensor Network of the corresponding quantum model. This is  the quantum Ising model on a square torus at its critical point obtained by   setting the parameters  of the Hamiltonian in Eq.  \ref{XY}  to $\gamma$=0 and $2 h=3.044$ \cite{mc-crit}. This approach is limited by the exponential growth of the Hilbert space with the system size. The benchmark calculations presented in Ref. \cite{tev} show that, nevertheless, it can be used safely at least  for tori of dimension up to $10 \times 10$. It has the strong advantage that  it allows us to extract the complete spectrum of the reduced density matrix $\rho_A$ for arbitrary blocks. It thus provides a way to directly compute $\tr \rho^{\alpha}$ for an arbitrary complex $\alpha$ and also  to directly compute the entanglement entropy. A detailed explanation on how to compute the reduced density matrices in this approach is contained in Ref.  \cite{tev}. We also postpone the discussion of the numerical  results for both the Tsallis and entanglement entropy extracted with this method to Section \ref{ent_ent}.

\resection{The two regions of Tsallis entropies }\label{sec:tay}
As we discussed in the introduction the Taylor expansions of  $\tr\rho_A^\alpha$ can be used to study the behaviour of the Tsallis and  R\'enyi entropies  as a function of  $\alpha$ and identify two very different regions.  In order to exemplify  this idea we have to express the Taylor expansion of   $\tr\rho_A^\alpha$ about  $\alpha=1$ in a convenient form. Let $p_i$ be the non 
vanishing eigenvalues of $\rho_A$. At first the normalisation of $\rho_A$ implies
\eq
\tr\rho_A=\sum_ip_i=1~.
\en
Through the chain of identities
\eq
\nn
\tr\rho_A^\alpha=\sum_ip_ip_i^{\alpha-1}=
\sum_ip_i\sum_{k=0}^\infty(\log\, p_i)^k\frac{(\alpha-1)^k}{k!}=
\sum_{k=0}^\infty\frac{\bra\,(\log \,p)^k\ket}{k!}(\alpha-1)^k
\en
\eq
=\sum_{k=0}^\infty\frac{\tr\rho_A (\log\rho_A)^k}{k!}\,(\alpha-1)^k
\equiv \sum_{k=0}^\infty c_k\,(\alpha-1)^k~,
\label{taylor}
\en
we define the Taylor coefficient $c_k =\frac{\tr\rho_A (\log\rho_A)^k}{k!} $. Since the eigenvalues of 
$\tr\rho_A$ are smaller than 1 the Taylor expansion is an alternating series,
i.e. $c_{k+1}=-c_k$. 
This is an important feature when discussing the possible location of the singularities in the complex $\alpha$ plane or in estimating the error we make 
in truncations of the Taylor expansions, as we shall see later.  
From  the spectrum of $\rho_A$ we can explicitly compute  the above Taylor 
expansion.

As an example, the method described in Ref. \cite{vlrk} can be used to compute the reduced density matrix of a block of spins  of an infinite critical Ising chain.  From  it,  we compute the moduli of the first Taylor  coefficients of  $\tr\rho_A^\alpha$  and  plot them in Fig.\ref{Figure:1} (a) for four different values of $L$.  It is interesting to notice that the plot shows a peak at some value of $k > 1$ and that its height  increases with $L$.

A similar behaviour can be observed in a quantum  two-dimensional critical 
Ising model on a torus $L\times L$ as shown in Fig.\ref{Figure:1} (b). There 
we plot again the absolute value of the Taylor coefficients of the expansion of 
$\tr\rho_A^\alpha$ about $\alpha=1$ for four different values of $L$. The 
subsystem $A$ in this case is half a torus. The reduced density matrices $\rho_A$ are obtained with the techniques described  in Ref. \cite{tev}.  By comparing the same expansion for different systems sizes, $L=4,6,8,10$ we appreciate how the peaks move with $L$ and, more importantly, their height increases with $L$.

  By varying the expansion centre of the Taylor expansions  $\alpha_o$  we can check the behaviour of the coefficients in other regions of the $\alpha$ axis. The  expression for the coefficients of the Taylor expansion about 
$\alpha=\alpha_o$ is
\eq
c_k[\alpha_o] =\frac{\tr\rho_A^{\alpha_o} (\log\rho_A)^k}{k!}~. 
\en 

We repeat the same study presented in Fig. \ref{Figure:1}  in the  region  $1 \le \alpha_o \le 2$ and observe that  the peaks of Fig. \ref{Figure:1} (a) and (b)  start moving  to left as we increase  the value of $\alpha_o$ to  eventually   disappear  at a threshold value $\alpha_o=n_c\le2$. In the region where the peaks are present we cannot obtain a good approximation of  $\tr\rho_A^\alpha$ by truncating the power series (\ref{taylor}) down to  only few terms. Also, the convergence of the series  (\ref{taylor}) deteriorates with  the size of the 
system, preventing the possibility  to find a good truncation scheme 
independent of the system's size. For this reason we call the region  $\alpha<n_c$ the non-perturbative region \footnote{ Starting from here, we will gradually drop the notation $\alpha_o$ in favour of $\alpha$. The reader should not be confused. Indeed the Taylor coefficient $c_k$ for $ \tr \rho_A^\alpha$ do not depend on $\alpha$ but rather on $\alpha_o$, the centre of the expansion. However
 $\alpha_o$ has the same support than $\alpha$ and we can safely simplify 
the notation.}.

For  $\alpha_o>n_c$ the coefficients  $c_k[\alpha_o]$ become 
monotone decreasing in modulus. As a consequence, in this region,  
the partial sum
$s_{k_o} = \sum_{k=0}^{k_o} c_k[\alpha_o](\alpha-\alpha_o)^k$ 
can be used to approximate $\tr\rho_A^\alpha$ with an error 
smaller than  $c_{k_o+1}[\alpha_o](\alpha-\alpha_o)^{k_o+1}$. 
In this case thus,  we  call  the  region  $\alpha>n_c$ the  perturbative region.
\begin{figure}[!hp]
\input Refig/ftr (a)
\input Refig/torustrans (b)
\caption{(a) Scaling behaviour of the threshold $n_c$ that separates the perturbative from the non-perturbative region,  as a function of the size $L$ of the block.  (a) 1D quantum critical Ising chain. The rhombi refer to  blocks 
embedded into an infinite chain, while the circles refer to  half of a finite chains of length 
$2L$. The threshold values  located  respectively at 
$n_i=1.414(5)$ and $n_h=1.391(1)$  are approached in both cases  with a power law with the same exponent $1/8$ . Data for the infinite chain are obtained  using the method of Ref. \cite{vlrk} while for  finite chains with the one dimensional version of the method of Ref. \cite{tev}.  (b)  2D quantum critical  Ising models on several  tori. The blocks are half tori. Here also the location of $n_c$ approaches its thermodynamic limit  as a power law with exponent $1/2$  and  the limit is located at $n_t=1.835(1)$. The data are obtained  with the technique introduced in Ref. \cite{tev}.}
\label{Figure:18}
\end{figure}

The threshold value $n_c$ is defined as the  minimal value of 
$\alpha$ such that
\eq
\vert c_{k+1}[\alpha]\vert\le\,\vert c_{k}[\alpha]\vert~,~~k=0,1,\dots~.
\label{che}
\en
that is, the Taylor coefficient $c_k$ are monotonically decreasing functions of $k$.
 In all the models we have considered, as soon as the criterion in Eq. \ref{che} is fulfilled by the second and third coefficients, that is,  
$\vert\,c_3[\alpha]\vert\le\vert\,c_2[\alpha]\vert$, it is also fulfilled by all the others pairs. Using this property   we can define $n_c$  as the solution of  Eq.(\ref{order}) anticipated in the introduction. We can also use 
\eq
s={\rm Sign}\left(\frac{{\rm d}^2\tr\rho_A^\alpha}{{\rm d}\alpha^2} + 
\frac13 \frac{{\rm d}^3 \tr\rho_A^{\alpha} }{{\rm d}\alpha^3}\right)~,
\label{osi}
\en
the sign of the expression (\ref{order})   defining $n_c$,
as a sort of order parameter distinguishing between the perturbative $(s>0)$
and non-perturbative $(s<0)$ `phases'. As expected, the value of the threshold depends on the size of the system, i.e. $n_c=n_c(L)$. By analysing this dependence we found that in our critical systems $n_c(L)$ obeys a simple power law
\eq
n_c(L)=n_o+b/L^\epsilon~,
\label{powl}
\en
where $\epsilon=\frac18$ in the 1D critical Ising model  and $\epsilon=\frac12$
in the 2D critical Ising model. It is also interesting to notice that  $n_o\le2$ for  all the   models we have analysed  
(see Figs. \ref{Figure:18} (a) and (b)). 

\resection{The nature of the transition}
\label{sing}
 Having uncovered  the presence of two different regions for  Tsallis (and  
R\'enyi)  entropies as  functions of $\alpha$, we now want to investigate 
whether these two regions are well separated phases divided by a 
transition point
or if instead they are different regions of the same phase smoothly connected through a cross-over. Of course, these two scenarios have  very different 
physical implications. In the case where the two regions are separated by a true phase transition for some real $n_c$ in the range $1<n_c<2$, there would be  no possibility to  analytically continue  the Tsallis entropy from one phase to the other. This, for example, would imply  that  the entanglement entropy 
could not be obtained via  the replica method.
This is unlikely to happen in  one dimensional critical systems since the entanglement entropy has been already successfully computed with the replica trick in some  1+1 CFT \cite{cardy-twist}. In two dimensions however, for the models we are considering, the entanglement entropy has not yet been obtained analytically and one can be concerned about  if with the replica trick is a viable method to perform such  a calculation.

In order to  settle the nature of the transition between the perturbative and non perturbative regions we start by  considering the radius of convergence of the Taylor expansion of $\tr\rho_A^\alpha$ around $\alpha_o=1$. 
It   indeed   measures the distance between the centre of the Taylor  expansion and the nearest singularity in the complex $\alpha$ plane. As the  expansions of $\tr\rho_A^\alpha$  is  an alternating series  its  possible singularity  belongs to the real axis  at the left side of the expansion centre.  We can also argue  on general grounds that the singularity should be at $\alpha$ greater or equal to zero.

 Indeed  we expect at least a singularity of  $\tr\rho_A^\alpha$ at $\alpha=0$, 
since  $\tr\rho_A^0$ measures the number of degrees of freedom of the 
subsystem $A$. This number  diverges exponentially with its size (Eq.(\ref{cc}), for example, presents an essential singularity at $\alpha=0$).

For the sake of completeness one should remember   that, when the system is in a product state (situation that does not arise close to a critical point)  there is no singularity 
at $\alpha=0$,  even in the thermodynamic limit. This happens, i.e.,  in the  
XY model described in (\ref{XY}) as one approaches the disorder line $h^2+\gamma^2=1$ in the ordered phase 
$(h\le1)$ . There the ground state 
becomes the superposition of two product states \cite{pe} and $\tr\rho^\alpha=2^{1-\alpha}$. 
Similarly,  there is a class of $z=2$ quantum critical points 
in two spatial dimensions  where a very simple form for $\tr\rho^\alpha_A $ 
has been proposed  \cite{fm}, namely
 $ \tr\rho^\alpha_A =(Z_D/Z_F)^{\alpha-1}$, where $Z_D$ and $Z_F$ are the 
partition functions of the associated system with suitable boundary 
conditions which select the subsystem $A$. 

In the cases of highly entangled states we consider here,  however, we do expect a singularity at $\alpha=0$ in the thermodynamic limit and therefore the  radius of convergence of the expansion  (\ref{taylor})  of  $\tr\rho_A^\alpha$  centred around   $\alpha_o=1$  should not exceed 1. A precise value for it  can be extracted using  the ratio test. This states that  if the limit $r=\lim_{k\to\infty}\left\vert\frac{c_k}{c_{k+1}}\right\vert$
exists, it coincides  with the convergence radius $r$. 
In a finite system, there is no true singularity  and 
the radius of convergence of the series is infinite. 
However we expect we are  still able to identify what would turn into a 
true singularity in the thermodynamic limit,  by studying the 
sequence $\left\{\left\vert\frac{c_k}{c_{k+1}}
\right\vert\right\}$. 

We anticipate that, notwithstanding there are clues to the presence of singularities in the thermodynamic limit, a careful analysis allows us to exclude it: the
only possible singularities of $\tr\rho^\alpha$ lie on the imaginary axis of 
$\alpha$.
 
\begin{figure}
\begin{center}
\includegraphics[width=0.7\linewidth]{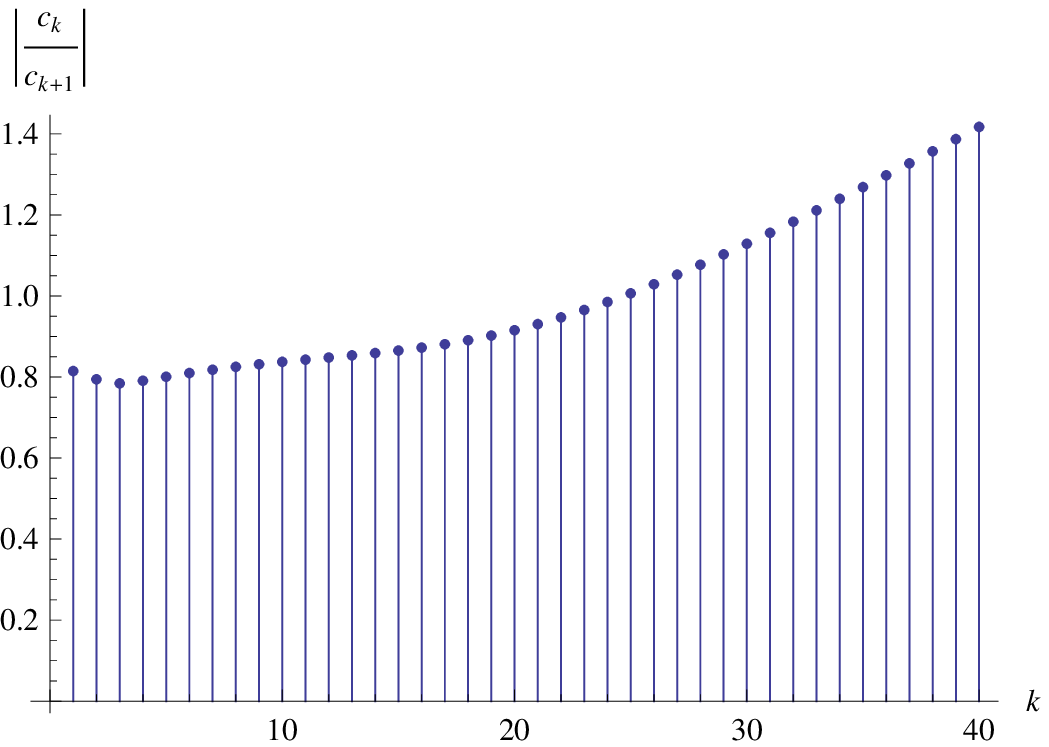} (a)\\
\includegraphics[width=0.7\linewidth]{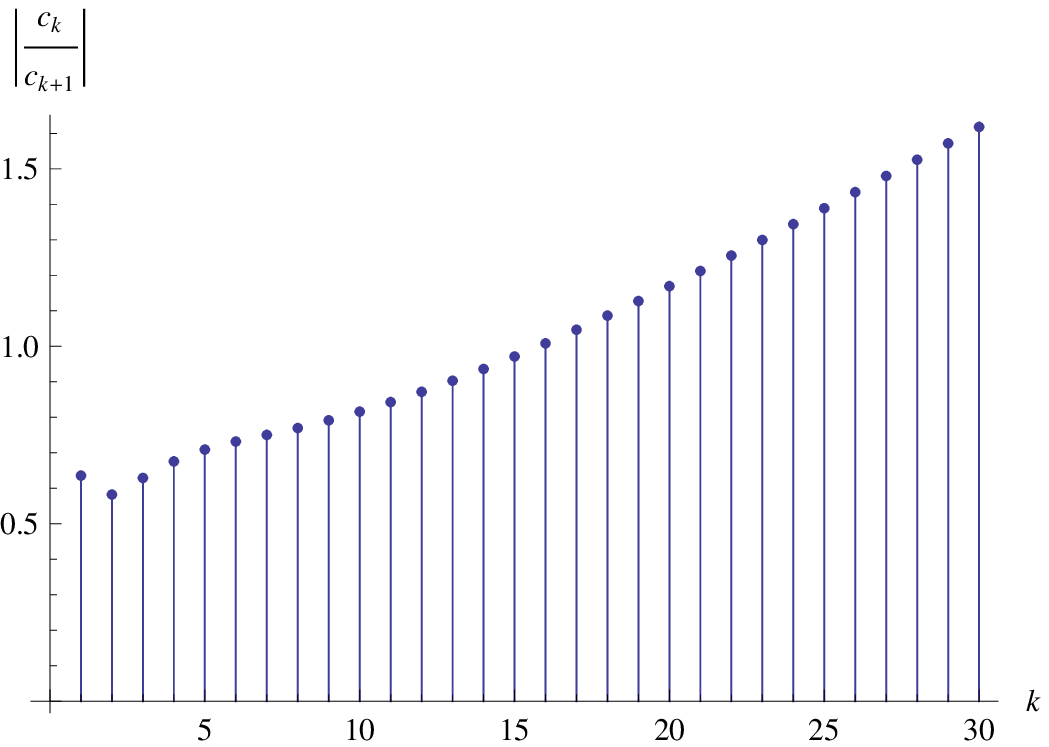} (b)
\end{center}
\caption{(a) Ratio $\frac{c_k}{c_{k+1}}$ as a function of $k$ obtained from  the Taylor coefficients of $\tr\rho_A^\alpha$ about $\alpha_o=1$   plotted in Fig. \ref{Figure:1}. The ratios oscillate for $k\le 20$  around $0.8$ 
before they run away. This could be a  hint of a finite radius of convergence
 for this Taylor series in the thermodynamic limit, generated by a conjectural singularity  located  somewhere around $1-0.8 =0.2$. In (b) the ratio test shows a similar behaviour  
for the $d=2$ critical Ising model on a torus $L\times L$ with $L=6$.}
\label{Figure:3}
\end{figure}

These ratio sequences show a sort of plateaux at a value  $r$ in a certain 
range of $k$  before they eventually  run  away for $k>k_o$.  
The point $k_o$ moves to the right within the size of the system, 
revealing longer and longer plateaux. 
This has been observed in all the models we have analysed. For instance, 
we may compute the required ratios  $\frac{c_k}{c_{k+1}}$ from the  Taylor coefficients  of a $L=512$ block of a critical one dimensional Ising model,  by using the coefficients plotted in Fig. \ref{Figure:1} (a). These ratios are  plotted in Fig. \ref{Figure:3}.

There we can see that the ratios oscillate for a while around  a 
value smaller  than 1 before running away. This suggests that a new 
singularity on the right side of  the one we already expect at $\alpha=0$  
might exist in the thermodynamic limit.

\begin{figure}
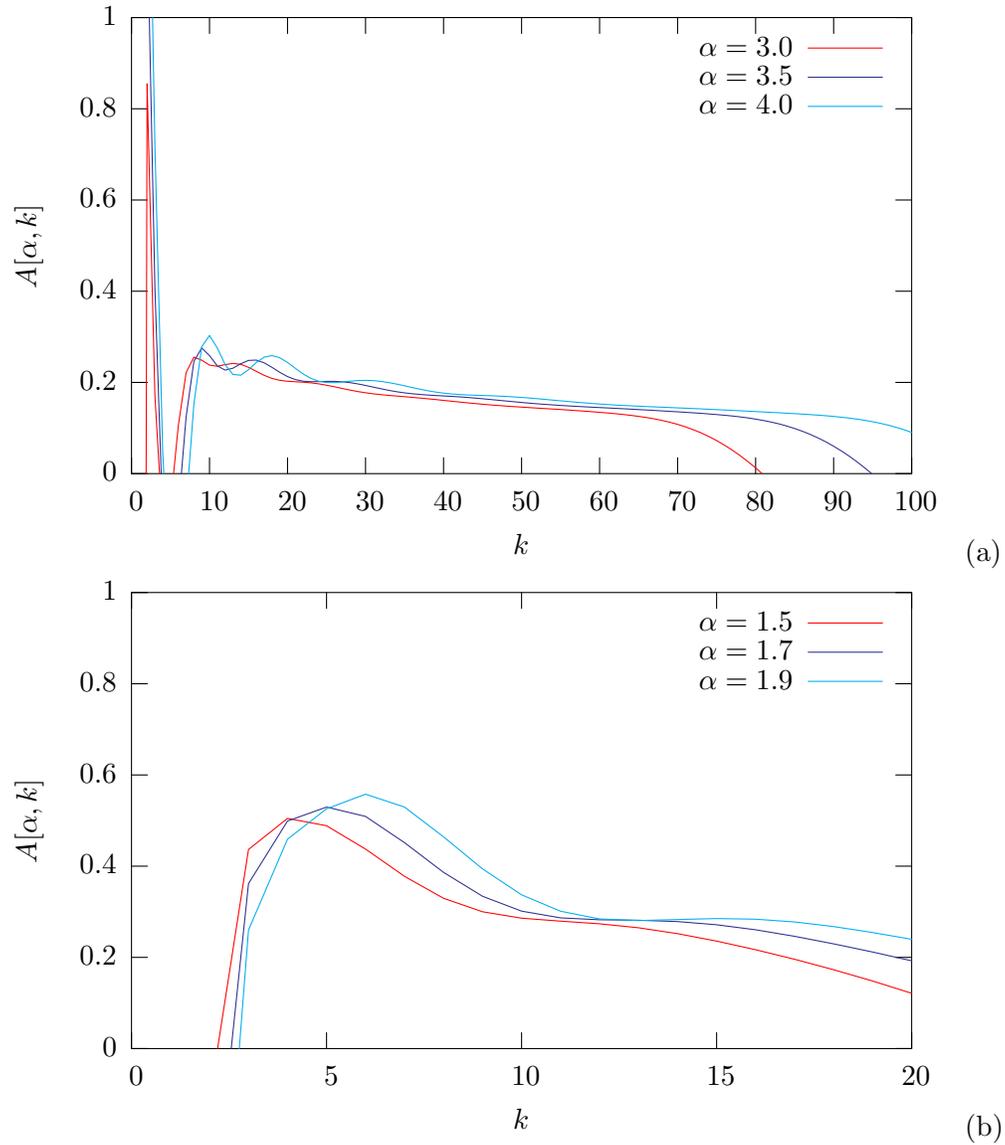

\input Refig/a1D \\
\input Refig/a2D
\caption{Plot of the function $A[\alpha,k]$ defined in (\ref{singa}) for different values of 
$\alpha$. This quantity should converge in the 
thermodynamic limit to the location of the alleged singularity. 
The system in (a) is the  same quantum Ising chain of Fig.\ref{Figure:1}.
The system in (b) is the critical Ising model on a torus $L\times L$ with $L=6$. Note that the position of the flexes on the $y$ axis is nearly independent 
of $\alpha$.}
\label{Figure:4}
\end{figure}

This scenario is further supported by studying  the Taylor expansions around other positive values of $\alpha$.
If there were  a true singularity on the real axis,  the function
\eq
A[\alpha,k]=\alpha-\left\vert\frac{c_k[\alpha]}{c_{k+1}[\alpha]}\right\vert~,
\label{singa}
\en
would converge to the coordinate location of the singularity in the 
thermodynamic limit  and would not dependent on the value of $\alpha$.
 In a finite system we expect that $A[\alpha,k]$  should approach to a 
constant value independent of $\alpha$  for a certain range of $k$, and this 
range should increase with the size of the system.  This behaviour  can be 
observed, for instance, in Fig. \ref{Figure:4} (a) where we plot $A[\alpha,k]$ 
for the $L=512$ block of the critical quantum Ising chain (the same already considered in Fig. \ref{Figure:1} (a)), and in Fig.\ref{Figure:4} (b) where we plot the same quantity for the two-dimensional Ising model on a torus of side $L=6$.

Still all the above signatures are not sufficient to conclude that there is a true singularity for real $\alpha$  in  the thermodynamic limit in 
 $\tr\rho^\alpha$.
As shown by Eq. (\ref{obs}), $\tr\rho_A^n$ can be written as the ratio of two 
partition functions.  This allows us to apply a Lee-Yang \cite{ly}  analysis to it. 

In discrete systems of 
finite size partition functions are analytic with respect to their parameters 
because they are finite summations of positive terms. For the same reason 
they do not have zeros on the real axis. However there can exist zeros on 
the complex plane of certain control parameter, $\alpha$ in our case, which are generally named Lee-Yang zeros \cite{ly}. In general, 
the Lee-Yang zeros are apart
from the real axis as long as the system size $L$ is finite. However, if a phase  transition occurs at a critical value $n_o\in\R$, the distribution of 
zeros becomes dense and accumulate on curves. The singularities of the free energy associated to this transition lie on the end points of these curves. 
As $L\to\infty$ these end points 
move towards the real axis,  eventually touching it at  the location of the critical point  $n_o$. 

\begin{figure}
\includegraphics[width=0.6\linewidth]{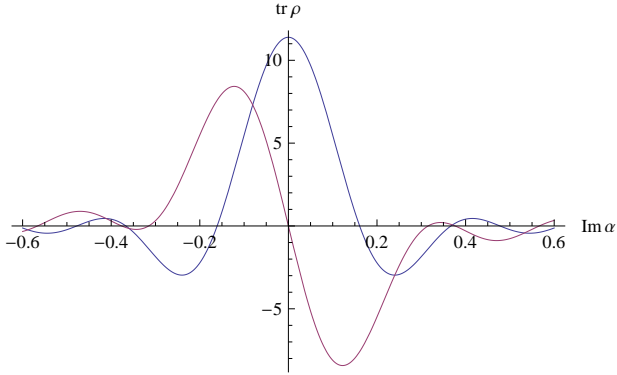}
\caption{Plot of the real and imaginary parts of $\tr \rho^\alpha$  
along the imaginary axis  $\Im\,m\,\alpha$ with $\Re e\,\alpha=0.3$ 
for the 2D Ising model on a $6 \times 6$ torus at the critical point. The imaginary part is the 
curve intersecting the origin. The intersection of the two curves identifies a pair of zeros at
$\Im\,m\,\alpha=\pm0.3655$.}
\label{Figure:14}
\end{figure}

In the case of   
$\tr\rho_A^n=\frac{Z_n(A)}{Z^n}$ besides the surviving zeros of the numerator
there could also be  poles due to zeros of the denominator. 
As an example, the exact  expression of the reduced density matrix in Eq. (\ref{rhoXY}) for the  non-critical XY chain,   presents  a singularity at $\alpha=0$  (the same as the singularity of (\ref{cc}))  but  no zeros. 
It contains instead  a huge number of double poles  at 
$\alpha=\pm i\frac{n\,\pi}{(2k-1)\epsilon}$ for any pair of integers
$k,n\in\N$. This set is dense on the imaginary $\alpha$ axis. The plot  of 
$\Re e\,\tr\rho_L^\alpha$ is presented in  Fig.  \ref{Figure:9} (a) and 
reflects this intricate structure.
A similar landscape can be observed by considering the quantum two dimensional critical Ising model on a torus (see Fig. \ref{Figure:9} (b)) where one can 
locate few zeros (see e. g. Figure \ref{Figure:14}) close to the region $\Re e\,\alpha=0.3$ where the ratio test give hints of a singularity. 

\begin{figure}[!ht]
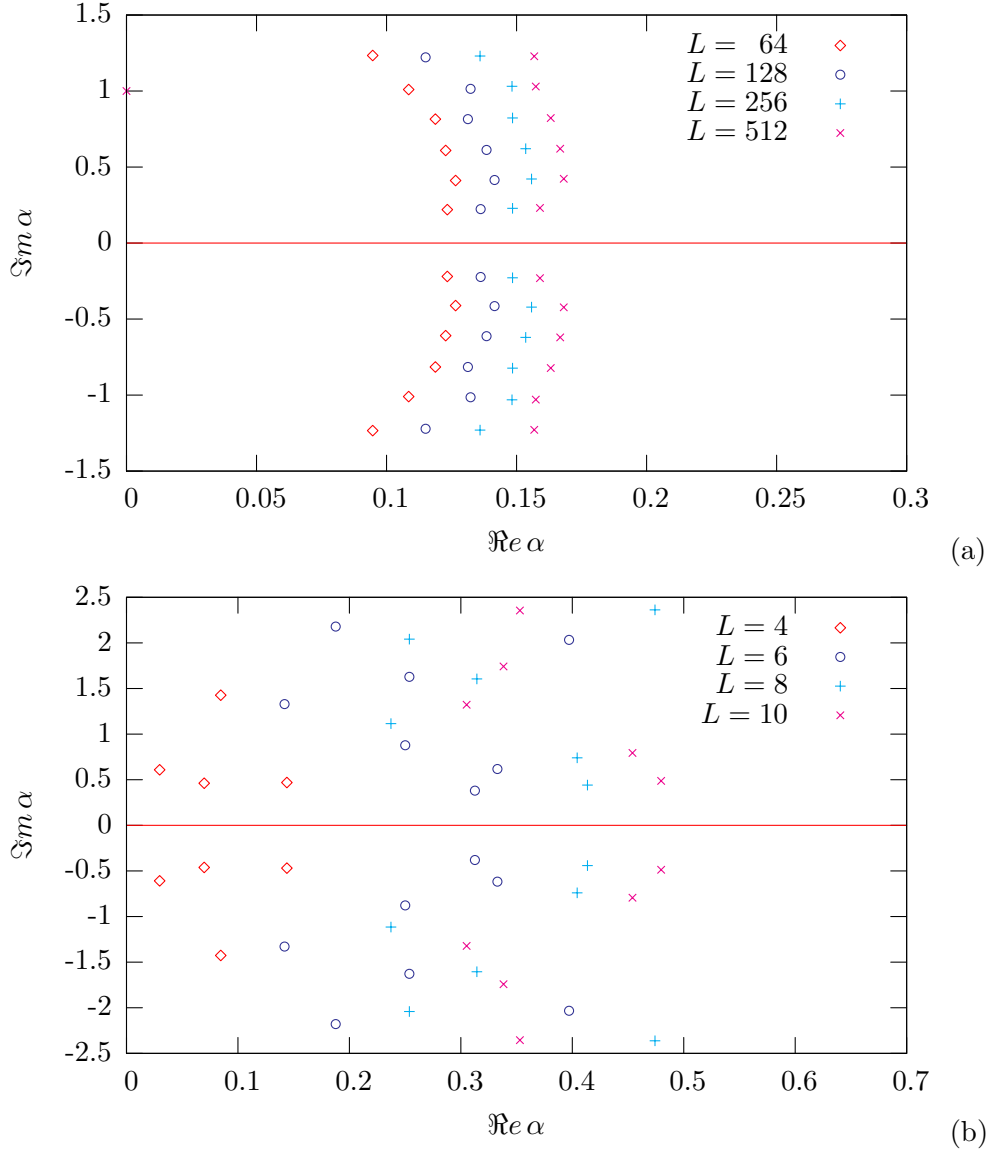

\input Refig/zeros (a)
\input Refig/zeros2D (b)
\caption{Finite size scaling of the zeros of $\tr\rho^\alpha_L$ in the complex $\alpha$-plane. 
for (a) several blocks of different length in the one dimensional XY model  and (b) several half tori of different size in  the 2D quantum Ising model at the critical point. In both cases,  the zeros do not 
approach the real axis nor their number grows with the size of the system.
This fact demonstrates that  they cannot originate a singularity for real $\alpha$ in  thermodynamic limit.}  
\label{Figure:145}
\end{figure}

The important outcome of this study is that  in all the models we have analysed both in $d=1$ and  in $d=2$, these Lee-Yang zeros, which  are true singular
 points of  the R\'enyi entropy (\ref{rets}) in the complex plane,   
do not become  denser as the size of the system increases  nor they approach 
the real axis (see Figure \ref{Figure:145}).  In the  thermodynamic limit 
then, the apparent singularity on the real axis identified by the ratio test of Taylor  coefficients does not occur.  The behaviour of  flexes 
described by  Eq.(\ref{singa}) is probably only a consequence of the 
presence of some structures in the complex $\alpha-$plane (zeros and/or peaks)
which do not evolve into a true singularity in the thermodynamic limit.      

The lesson we learn from this analysis is twofold. On the one hand there is 
no true phase transition  between the two regions where  the Tsallis entropies 
show very different behaviour. The transition is only a smooth cross-over. 
On the other hand the behaviour of $\tr\rho^\alpha$ shows an intricate landscape
in the region $\Re e\,\alpha<n_c$ while it is very smooth in the complementary region $\Re e\, \alpha>n_c$. As a consequence a naive analytical continuation 
of the results obtained for the Tsallis entropies in the perturbative 
region is likely to fail to reproduce the correct entanglement entropy. 
We will provide an example of this fact in the following section.  
\begin{figure}
\begin{center}
\input Refig/surexact (a)
\input Refig/sur6x6 (b)
\caption{ Plot of $\Re e\,\tr\rho^\alpha_L$ for the non-critical 
XY chain in the  complex replica plane (a) and  in the 2D Ising model at
 at  the critical point for half a $6\times 6$  torus (b). Both  plots show a 
very rugged landscape  with several dips and peaks.}
\label{Figure:9}
\end{center}
\end{figure}

 As an aside, it is  interesting to observe the  striking resemblance 
between the location and shape of the zeros 
distribution of Fig. \ref{Figure:145} and those found in certain disordered 
systems \cite{ok}. In the latter case the replicas are coupled together no longer by the subsystem $A$, but by the quenched average, which restores the translational invariance of the system. As a consequence the latter system develops 
in thermodynamic limit a true singularity associated to  replica 
symmetry breaking.  

\begin{figure}
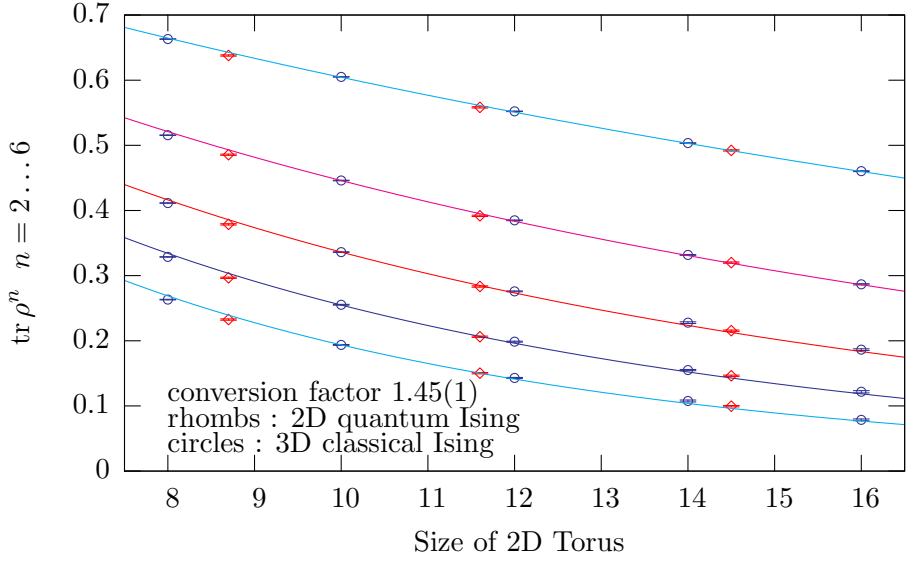

\input Refig/conversion
\caption{Fit of the Monte Carlo estimations of $\tr\rho_L^n$ (rhombi) to the 
function (\ref{fit}) in a 3D critical Ising model. The circles are 
the values calculated in the corresponding quantum system in 2+1 dimensions
 with the method of \cite{tev} with the length rescaled by a suitable conversion factor.}
\label{Figure:0}
\end{figure}

\resection{The entanglement entropy mismatch}\label{ent_ent}
In order to test the consequences of the intricate structure of $\tr \rho^\alpha$ in the complex plane we have performed a numerical experiment. This consists in computing the entanglement entropy using the replica trick from the results obtained in a Monte Carlo simulation. For the 2D Ising model on a torus indeed there are still no analytical predictions on value of the parameter appearing in the proposed form of the scaling of the entanglement entropy. Some numerical results have been recently obtained in Ref. \cite{tev}. Here we have computed  $\tr \rho_A^n$ for $n=2,3,4,5,6$  with the Monte Carlo technique described in Sect. \ref{mc} when  $A$ is a half of the torus, the same geometry studied in Ref. \cite{tev}. The results we  obtain agree with the ones computed from the data of  Ref. \cite{tev} provided we introduce a suitable conversion factor between the lattice spacings of these two different regularisation schemes.
 A good agreement  is indeed  obtained  by multiplying the  lengths measured in the Hamiltonian approach by a
conversion factor $f\simeq1.45$ (see Fig.\ref{Figure:0}). 
This agreement  is a good check of the validity of both methods. To our knowledge it is  also the first direct comparison between two different non-perturbative techniques on entanglement entropy, one involving quantum mechanical tensor 
network simulations  and the other one  using statistical mechanics  
Monte Carlo integration. This is what we expect from universality, indeed, 
these two methods deal with very different systems (in one case the model 
studied is the 3D classical  Ising model while in the other case it is the 
2D Quantum Ising model). They show the same behaviour close to the 
critical point because they belong to the same universality class.

We can now try to extract a function  $f(L,n)$  from  the data  describing the behaviour of  $\tr\rho^n$ in the whole range of $n$  and $L$ considered . The function should depend  on only few  parameters  and is  chosen in such a way to
 agree, in the $n\to1$ limit, with  the  expected 
general form of the critical entanglement entropy in 2+1 dimensions 
\cite{ch,sach}. A sufficiently good fit (a reduced $\chi^2= 1.75$ 
for 17 degrees of freedom) is given by the three-parameter function
\eq
f(n,L)=\exp[-a\,(n-\frac1{n^3})2\,L-(b+\frac c{2L})(n-\frac1n)]
\label{fit}
\en
with
\eq
a= 0.01177(92)\,, ~b=0.0594(55)\,,~c=-0.378(73)~.
\label{num_est}
\en          




The `obvious' extension to real $n$, is performed by  assuming that 
this functional form should  also be valid for real $n$. The entanglement entropy then is given by  
\eq
S_L\equiv -\tr\rho_A\log \rho_A \, =^{\hskip -.2 cm ?}~-\lim_{n\to1}
\frac{{\rm d}}{{\rm d}n}f(n,L)=8\,a\,L+2b+c/L.
\en 
Due to our choice of the parametrisation (\ref{fit}) the entanglement entropy has the expected functional form $S_L=c_1\,2L+c_0+
\frac{c_{-1}}{2L}.$ The numerical coefficients we extract however, apart from the 'area term' $c_1$, do not coincide with the ones extracted from the direct study of the entanglement entropy computed in Ref. \cite{tev}. If we call the  parameter extracted in  Ref. \cite{tev}
 $S_L=c_1' 2L+c_0'+\frac{c_{-1}'}{2L}$, taking into account the conversion factor $f$, we would  expect from the values reported in (\ref{num_est}) that 
\eq
c_1'=c_1\,f=0.0682(5)~,\,c_0'=c_0=0.12(1)~,\,c_{-1}'=c_{-1}/f=-0.52(10).
\en
 On the other hand their direct numerical determination  \cite{tev} provides  different results
\eq
c_1'=0.06722(18)~,\,c_0'=0.0250(21)~,\,c_{-1}'\simeq0~.
\en

\begin{figure}
\includegraphics[width=0.8\linewidth]{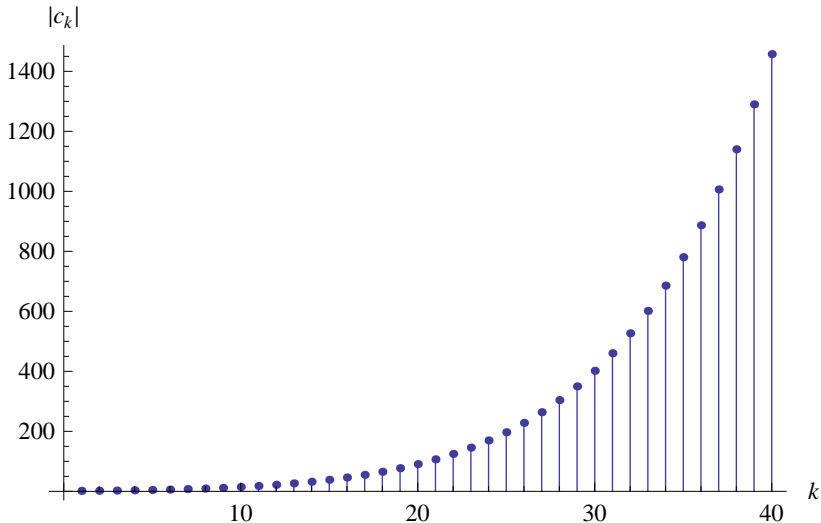}
\caption{Plot of the moduli of the coefficients of the Taylor expansion 
about $n=1$ of Eq.(\ref{cc}) with $r_n=1$. Although it fits perfecltly 
the behaviour of $\tr\rho^n_L$ of the same system of Fig. \ref{Figure:1} (a) 
with $L=512$ in the range $n\ge2$, its analytic 
continuation to $n\simeq 1$ region does not reproduce the true Taylor 
expansion plotted in Fig.\ref{Figure:1} (a).}
\label{Figure:5}
\end{figure}

We attribute the mismatch of  the entanglement entropy to  the 
the different behaviour of the the  regions $\Re e\,\alpha<n_c$ and 
$\Re e\,\alpha>n_c$. 

A further confirmation of this can be obtained by computing, in  the 
Hamiltonian approach,
$\tr\rho_A^\alpha$ for some  real (or complex) $\alpha$ in the non-perturbative region.  If we add to the fitted data some of these points  
the quality of the fit to (\ref{fit}) rapidly deteriorates.

One can check that a 
similar phenomenon holds also for $d=1$ models. Indeed, considering  again the 
critical XY   system in which the subsystem $A$ is composed by a block of  
$L=512$ adjacent spins, $\tr\rho_L^n$ can be evaluated with  the method of \cite{vlrk} and the data with $n\ge2$ can accurately fitted with
Eq.(\ref{cc}) ignoring the $n$ dependence of the non-universal coefficient $r_n$ which is known to be neglegible in such a range \cite{cl}.  If now we assume  that this functional form with $r_n=1$ is also valid for real $n$ 
and try to evaluate this quantity near $n=1$ we obtain incorrect results for the entanglement entropy. As an example  the Taylor coefficients of this fitting function are plotted in 
Fig. \ref{Figure:5} and should be compared  with the  same coefficients directly evaluated on 
 $\tr\rho_L^n$ (see Fig.\ref{Figure:1} (a) ). They  show a completely different shape.  Again this mismatch has to be attributed to  having 
applied to the fitting function (\ref{cc}) with $r_n=1$, which accurately 
describes the behaviour of $\tr\rho^n$ in the perturbative region 
(i.e. $n>n_c$), the  `obvious' extension to real $n$  close to $n=1$ which 
lies in the non-perturbative region.

\resection{Discussion}
\label{sec:con}
In this work we analysed the trace of the $n-$th power 
of the reduced density matrix $\tr\rho_A^n$ associated to a subsystem $A$
in some Ising-like quantum models in both  one and two spatial dimensions 
near and at a quantum critical point at zero temperature. We  unveiled 
that, depending on  the value of $n$, $\tr \rho_A^n$  shows two  very 
different  behaviours. If $n$ is larger than a threshold 
value $n_c$ -located in  the interval $1<n<2$- its behaviour is very smooth. 
In particular the alternating coefficients of its Taylor expansions, $c_k$ are 
monotone decreasing. In this regime  the error involved  in a truncation of 
the series does not exceed the first 
excluded term and can be always chosen to be very small. We label this region `perturbative' region. 

In the range $n<n_c$, on the other hand,   the $\vert c_k\vert$'s are no longer monotone decreasing function of $k$  but they present a characteristic  peak (see Fig.\ref{Figure:1} ) whose height increases with the size of the block considered. Any truncated Taylor expansion, in this region,  
provides  an increasingly  poor approximation of $\tr\rho_A^n$   as the size of $A$ increases. This is the region that we label `non-perturbative'.

The transition between these two regions turns out to be a smooth cross-over. 
The dependence of the threshold $n_c(L)$ on the size of the subsystem obeys 
a power law described in Eq. (\ref{powl}). We  determined the  critical 
exponents ruling this power law for the Ising model in both one and two 
dimensions. It would be interesting to investigate whether this 
cross-over is somehow related to the phase transition close to n=2 
recently observed  in \cite{sach} near $d=3$.

The above scenario has important consequences when considering the analytical 
continuation involved in the computation  of the entanglement entropy with 
the replica trick. This  continuation  is possible  
in one and two dimensions, as there is no phase transition between the 
perturbative region 
(where both analytical results from QFT and numerical results from Monte 
Carlo algorithms are available)  and the non-perturbative region 
(where the entanglement entropy is located). However its form is not 
the naive expression obtained by 
promoting  the simple expression obtained in the perturbative region 
for integer $n$  to general $\alpha$, as we have  explicitly shown in  
Section \ref{ent_ent} .

From our analysis, the R\'enyi and Tsallis entropies seem to emerge as a
 privileged measures of entanglement in QFT.  Although they lack the nice 
theoretical information properties  of the entanglement entropy \cite{ente},  
their analytic properties as functions of the $\alpha$ parameter seem to 
contain some new information  on the entanglement. This has been 
considered recently in the context of quantum criticality 
\cite{pasq,frad,sach} and  topological order \cite{wen}.

LT acknowledges the financial support from the QuSim group of the University of Queensland, the stimulating discussions with P. Corboz  and the enlightening comments on entanglement measures of Prof. G. Vidal.

%
%

\end{document}